\documentclass[aps,prx,twocolumn]{revtex4-2}
\usepackage{graphicx,amsmath,amsfonts,amssymb}% Include figure files
\usepackage{bm}% bold math
\usepackage[mathlines]{lineno}% Enable numbering of text and display math
\usepackage[normalem]{ulem} %to strike the

\usepackage[dvipsnames]{xcolor}
\usepackage{ulem}
\usepackage{wasysym}
\usepackage{capt-of}

\usepackage{soul}
\global\long\def\abs#1{\left|#1\right|}

\usepackage{xcolor,hyperref}
\hypersetup{
   colorlinks,
   linkcolor={blue!50!black},%{red!80!black},
   citecolor={blue!50!black},
   urlcolor={blue!80!black}
}
\begin{document}

\title{Unsteady slip pulses under spatially-varying prestress}
\author{Anna Pomyalov}
\author{Eran Bouchbinder}
\email{eran.bouchbinder@weizmann.ac.il}
\affiliation{Chemical and Biological Physics Department, Weizmann Institute of Science, Rehovot 7610001, Israel}

\begin{abstract}
It has been recently established that self-healing slip pulses under uniform background/ambient stress (prestress) $\tau_{\rm b}$ are intrinsically unstable frictional rupture modes, i.e., they either slowly expand or decay with time $t$. Furthermore, their spatiotemporal dynamics have been shown to follow a reduced-dimensionality description corresponding to a special $L(c)$ line in a plane defined by the pulse propagation velocity $c(t)$ and size $L(t)$. Yet, uniform prestress is rather the exception than the rule in natural faults. Here, we study the effects of a spatially-varying prestress $\tau_{\rm b}(x)$ (in the fault direction $x$) on 2D slip pulses, initially generated under a uniform $\tau_{\rm b}$ along a rate-and-state friction fault. We consider both periodic and constant-gradient prestress distributions $\tau_{\rm b}(x)$ around the reference uniform $\tau_{\rm b}$. For a periodic $\tau_{\rm b}(x)$, pulses either sustain and form quasi-limit cycles in the $L\!-\!c$ plane or decay predominantly monotonically along the $L(c)$ line, depending on the instability index of the initial pulse and the properties of the periodic $\tau_{\rm b}(x)$. For a constant-gradient $\tau_{\rm b}(x)$, expanding and decaying pulses closely follow the $L(c)$ line, with small systematic shifts determined by the sign and magnitude of the gradient. We also find that a spatially-varying $\tau_{\rm b}(x)$ can revert the expanding/decaying nature of the initial reference pulse. Finally, we show that a constant-gradient $\tau_{\rm b}(x)$, of sufficient magnitude and specific sign, can lead to the nucleation of a back-propagating rupture at the healing tail of the initial pulse, generating a bilateral crack-like rupture. This pulse-to-crack transition, along with the above-described effects, demonstrate that rather rich rupture dynamics can emerge from a simple, spatially-varying prestress. Furthermore, our results show that as long as pulses exist, their spatiotemporal dynamics are related to the special $L(c)$ line, providing an effective, reduced-dimensionality description of unsteady slip pulses under spatially-varying prestress.
\end{abstract}
\maketitle

\section{B\lowercase{ackground and motivation}}
\label{sec:intro}

Classical models of earthquakes, which consider a single frictional rupture that propagates along a homogeneous and planar fault separating unbounded linear elastic blocks in the presence of a uniform background/ambient stress (prestress) $\tau_{\rm b}$, offer insight into the physics and mechanics of faulting~\cite{Rice1980a,Freund1998,Scholz2002}. Single frictional rupture is classified into crack-like modes and self-healing slip pulses~\cite{Heaton1990,Perrin1995,lu2007pulse,Gabriel2012,Brener2018}, where the latter feature a characteristic size $L$ that implies a finite slip duration at points along the fault. Yet, our understanding of even these relatively simple models is incomplete.

Major recent progress has been made in understanding the spatiotemporal dynamics of self-healing slip pulses propagating along rate-and-state friction faults under a uniform prestress $\tau_{\rm b}$~\cite{pomyalov2023dynamics}. It has been shown that slip pulses are intrinsically unstable objects that cannot propagate steadily, i.e., they either expand ($\dot{L}(t)\!>\!0$) or decay ($\dot{L}(t)\!<\!0$) with increasing time $t$. Yet, it has been shown that the unsteady pulses dynamics are slow and follow a special $L(c)$ line in a plane defined by the pulse propagation velocity $c(t)$ and size $L(t)$. The special $L(c)$ line corresponds to steady pulse solutions parameterized by $\tau_{\rm b}$, which is composed of unstable points that nevertheless form a dynamic attractor at a given $\tau_{\rm b}$. As such, special $L(c)$ line should be regarded as an equation of motion for unsteady pulses under uniform prestress. This reduced-dimensionality description remains valid for other pairs of pulse quantities, e.g., the peak slip velocity $v_{\rm p}(t)$ and $L(t)$~\cite{pomyalov2023dynamics}.

Real faults in the earth crust generally feature spatially-varying prestress~\cite{hanks1974faulting,kanamori1978seismological,hartzell1979horse,aki1979characterization,bache1980source,beroza1996short,matsumoto2018prestate,mildon2019coulomb}. Such nonuniform prestress distributions may emerge from previous inhomogeneous slip histories, or to effectively mimic either spatial variations in the frictional strength or geometrical fault irregularities. The latter include barriers~\cite{aki1979characterization}, bends~\cite{mildon2019coulomb}, double fault bends~\cite{lozos2011effects}, branched fault systems~\cite{duan2007nonuniform} and fault surface roughness~\cite{Fang2013,heimisson2020crack}. Consequently, it has been recognized quite a long time ago that earthquake models should be extended to include spatially-varying prestress distributions~\cite{aki1979characterization,day1982three}. It is particularly interesting to understand whether and to what extent a nonuniform prestress may contribute to the emergence of slip complexity, including phenomena such as non-self-healing slip pulses~\cite{beroza1996short}, rupture arrest, back-propagating rupture~\cite{idini2020fault,ding2024back} and pulse-to-crack transitions~\cite{Perrin1995,lu2007pulse,Gabriel2012,Brener2018,brantut2019stability,heimisson2020crack}.

Indeed, some works have studied the effects of nonuniform prestress on frictional rupture dynamics in various spatial dimensionalities, including in three dimensions (3D)~\cite{day1982three}, in two dimensions (2D)~\cite{johnson1990initiation,elbanna2011pulselike} and in one dimensional (1D) space~\cite{barras2023earthquakes}, to list a few among others. The PhD thesis of~\cite{elbanna2011pulselike} (see chapter 5 therein) focused on 2D pulse propagation in nonuniform prestress fields, arguing that the interaction of pulses with inhomogeneous prestress distributions is of particular interest as the finite size of pulses implies larger sensitivity to prestress spatial variations compared to crack-like rupture, which may tend to smooth out their effect.

Our goal in this paper is to study 2D unsteady pulse dynamics under spatially-varying prestress distributions $\tau_{\rm b}(x)$ of two classes. One is a periodic prestress and the other is a constant-gradient prestress. We will employ the above-mentioned reduced-dimensionality description of unsteady slip pulses, originally developed under uniform prestress~\cite{pomyalov2023dynamics}, to quantify pulse dynamics under nonuniform $\tau_{\rm b}(x)$ and to test the degree by which this theoretical framework extends to a broader class of problems. Our analysis also employs well defined initial pulse states, corresponding to controllably perturbed steady-state pulse solutions.

\section{P\lowercase{roblem setup}: I\lowercase{nitial pulses and the spatially-varying prestress} $\tau_{\rm b}(x)$}
\label{sec:setup}

We consider slip dynamics along a rate-and-state friction fault separating two infinite linear elastic blocks under anti-plane shear. The displacement field is $u_z(x,y,t)$, where $y\!=\!0$ defines the fault and $z$ is perpendicular to the $xy$ plane. The frictional strength $\tau(v,\phi)$ is a functional of the slip velocity field $v(x,t)\!\equiv\!2\partial_t u_z(x,y\!=\!0^{+},t)$ and of the state field $\phi(x,t)$. The latter follows the ``aging law'' $\partial_t\phi(x,t)\!=\!1\!-\!v(x,t)\,\phi(x,t)/D$, where $D$ is a characteristic slip displacement~\cite{Dieterich1979a,Ruina1983,Nakatani2001,Marone1998a,Baumberger2006Solid}. $\tau(v,\phi)$, whose explicit expression is given in the Supplementary Material (SM, appended below), features an $N$ shape under steady-state sliding (for which $\phi\!=\!D/v$), with a transition from rate-weakening to rate-strengthening friction occurring at a local minimum $(v_{\rm min},\tau_{\rm min})$ of the steady-state friction strength (see SM).

Spatiotemporal slip dynamics for $t\!>\!0$ are described by the boundary integral equation~\cite{das1980numerical}
\begin{equation}
\label{eq:BIM}
\tau[v(x,t),\phi(x,t)] = \tau_{\rm b}(x) -\frac{\mu}{2c_{\rm s}}v(x,t) + s(x,t) \ ,
\end{equation}
balancing the frictional strength on the left-hand-side with the interfacial shear stress on the right-hand-side. The latter features three contributions; the first is the spatially-varying prestress $\tau_{\rm b}(x)$, the second is the so-called radiation-damping term (where $\mu$ is the shear modulus and $c_{\rm s}$ is the shear wave-speed) and the third, $s(x,t)$, is a spatiotemporal convolutional integral that accounts for the long-range interaction of different parts of the fault, mediated by bulk elastodynamic deformation. Equation~\eqref{eq:BIM}, along with the $\phi$ evolution equation, are solved in the spectral domain~\cite{Breitenfeld1998} using the open-source library $cRacklet$~\cite{roch2022cracklet} in a periodic domain of size $W$, i.e., $x\!\in\![-\tfrac{1}{2}W,\,\tfrac{1}{2}W]$ with periodic boundary conditions (see SM). The problem formulation is completed once the initial conditions and $\tau_{\rm b}(x)$ are specified.
%%%%%%%%%%%%%%%%%%%%%%%%%%%%%%%%%%%%%%%
\begin{figure}[ht!]
\center
\includegraphics[width=0.49\textwidth]{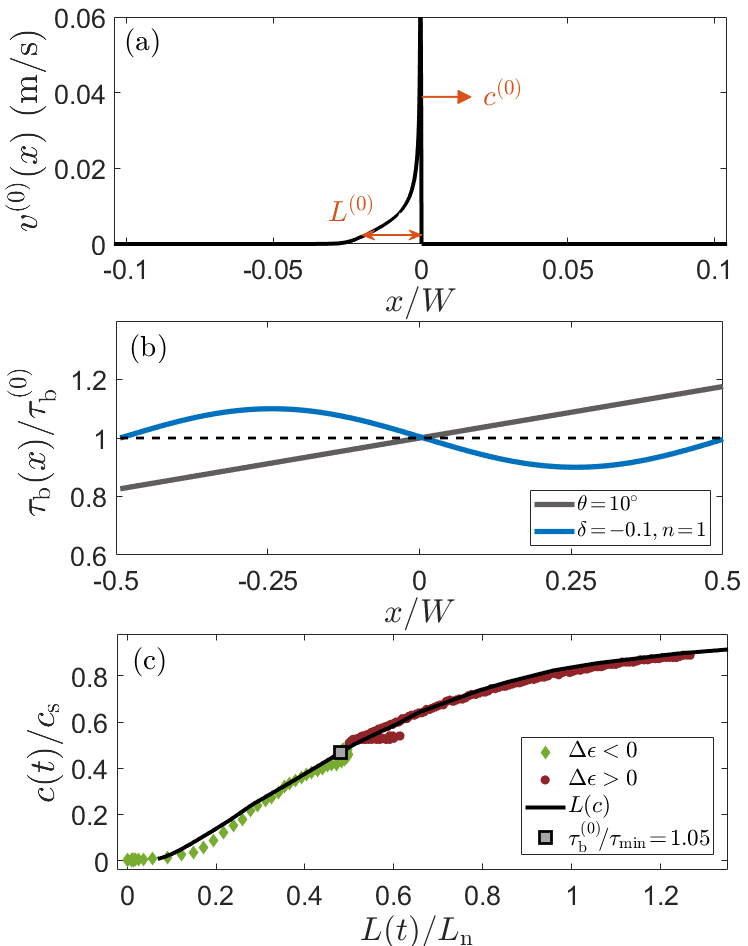}
\vspace{-0.1cm}
\caption{(a) The slip velocity field $v^{\mbox{\tiny{(0)}}}(x)$ of a steady-state slip pulse characterized by size $L^{\mbox{\tiny{(0)}}}$ (marked) and propagation velocity $c^{\mbox{\tiny{(0)}}}$ (from left to right, marked). $x\!=\!0$ corresponds to the location of the peak slip velocity $v_{\rm p}$ (not marked). (b) Examples of members of the two nonuniform prestress $\tau_{\rm b}(x)$ classes discussed in the text (see legend for the parameters). Note that while $x\!=\!0$ here is aligned with $x\!=\!0$ of panel (a), there is an order of magnitude difference in the $x$ axis range of the two panels. (c) Unsteady slip pulse dynamics under uniform prestress $\tau^{\mbox{\tiny{(0)}}}_{\rm b}/\tau_{\rm min}\!=\!1.05$, presented in the $L\!-\!c$ plane ($L_{\rm n}$ is a normalization length, see SM). The special line $L(c)$ is marked by the solid line~\cite{pomyalov2023dynamics}. The steady-state pulse corresponding to $\tau^{\mbox{\tiny{(0)}}}_{\rm b}/\tau_{\rm min}\!=\!1.05$, marked by the square, is initially perturbed with $\Delta\epsilon\!>\!0$ (brown circles) or with $\Delta\epsilon\!<\!0$ (green diamonds), and their subsequent evolution ($t\!>\!0$) is plotted in the $L\!-\!c$ plane. As extensively discussed in~\cite{pomyalov2023dynamics}, unsteady pulse dynamics closely follow the steady-state $L(c)$, whether the pulse expands ($\Delta\epsilon\!>\!0$) or decays ($\Delta\epsilon\!<\!0$).}
\label{fig:fig1}
\vspace{-0.25cm}
\end{figure}
%%%%%%%%%%%%%%%%%%%%%%%%%%%%%%%%%%%%%%%

In setting up the initial conditions, our goal is to employ well defined and controlled states that will allow us to address our main question regarding the interaction of slip pulses with nonuniform prestress distributions. Consequently, we are not interested in spontaneous frictional rupture nucleation under nonuniform prestress, but rather in generating propagating self-healing slip pulses with known properties. To this end, we invoke the steady-state pulse solutions of~\cite{pomyalov2023self}, obtained under a uniform prestress, to be denoted as $\tau^{\mbox{\tiny{(0)}}}_{\rm b}$. Each such solution is characterized by time-independent slip velocity $v^{\mbox{\tiny{(0)}}}(x)$ and state $\phi^{\mbox{\tiny{(0)}}}(x)$ fields, which propagate steadily at a velocity $c^{\mbox{\tiny{(0)}}}$ and feature a characteristic size $L^{\mbox{\tiny{(0)}}}$, see Fig.~\ref{fig:fig1}a. Such solutions were shown to correspond to ``saddle configurations'' separating expanding from decaying pulses. This implies that small positive perturbations of the steady-state pulse, which we quantify through an instability index $\Delta\epsilon\!>\!0$ (see SM), lead to a growing pulse, while small negative perturbations of the steady-state pulse with $\Delta\epsilon\!<\!0$ (see SM), lead to a decaying pulse, where $|\Delta\epsilon|$ determines the initial rate of instability.

Yet another major advantage of using controllably perturbed steady-state pulses, obtained under uniform prestress $\tau^{\mbox{\tiny{(0)}}}_{\rm b}$, as initial conditions for Eq.~\eqref{eq:BIM}, is that they set a dynamical reference for the analysis under spatially-varying prestress $\tau_{\rm b}(x)$ to follow. Specifically, as already stated, it has been recently demonstrated that the unsteady pulse motion under uniform prestress $\tau^{\mbox{\tiny{(0)}}}_{\rm b}$, e.g., as quantified by the time evolution of $L(t)$ and $c(t)$, closely follows the special $L(c)$ line in the $L-c$ plane, whether the pulse grows/expands ($\dot{L}(t)\!>\!0$ for $\Delta\epsilon\!>\!0$) or decays ($\dot{L}(t)\!<\!0$ for $\Delta\epsilon\!<\!0$), see Fig.~\ref{fig:fig1}c. The special $L(c)$ line, which corresponds to steady-state pulse solutions parameterized by $\tau^{\mbox{\tiny{(0)}}}_{\rm b}$~\cite{pomyalov2023self}, has been shown to be a dynamic attractor (also for pulses nucleated away from it) and hence to provide a reduced-dimensionality description of unsteady slip pulses under uniform prestress~\cite{pomyalov2023dynamics}. The degree by which this reduced-dimensionality description extends to unsteady pulse dynamics under spatially-varying prestress is a central question addressed in this work.

To complete the problem formulation, we need to specify the spatially-varying prestress distributions $\tau_{\rm b}(x)$ in Eq.~\eqref{eq:BIM}. A first class of prestress distributions corresponds to a periodic (sinusoidal) variation of the form
\begin{equation}
    \quad\tau_{\rm b}(x) = \tau^{\mbox{\tiny{(0)}}}_{\rm b}\left[1+\delta\sin(2\pi\,n\,x/W) \right] \ .
\label{eq:sin}
\end{equation}
Here, the prestress distribution is characterized by two parameters: a dimensionless amplitude $\delta$ and a wavelength $\lambda\!=\!W/n$ determined by $n$ (expressed in terms of the domain size $W$). A second class of prestress distributions corresponds to a constant spatial gradient $d\tau_{\rm b}(x)/dx$ that is characterized by a tilt angle $\theta$ according to
\begin{equation}
    \quad\tau_{\rm b}(x) = \tau^{\mbox{\tiny{(0)}}}_{\rm b}\left[1+2\tan(\theta)\,x/W \right] \ ,
\label{eq:tilt}
\end{equation}
where $\theta\!>\!0$ corresponds to a counterclockwise tilt. Examples of members of the two $\tau_{\rm b}(x)$ classes are shown in Fig.~\ref{fig:fig1}b.

The fault average of both distributions in Eqs.~\eqref{eq:sin}-\eqref{eq:tilt} satisfies $\langle\tau_{\rm b}(x)\rangle_x\!=\!\tau^{\mbox{\tiny{(0)}}}_{\rm b}$, where the latter characterizes the uniform prestress for which the perturbed steady-state (serving as the $t\!=\!0$ initial condition) was obtained. This setup mimics physical situations in which an unsteady slip pulse obtained under a uniform prestress $\tau^{\mbox{\tiny{(0)}}}_{\rm b}$ enters at $t\!=\!0^+$ a region of nonuniform prestress $\tau_{\rm b}(x)$ (with a fault average that equals $\tau^{\mbox{\tiny{(0)}}}_{\rm b}$), and its subsequent $t\!>\!0$ dynamics are tracked. Note also that we are interested in prestress distributions that vary rather slowly on the characteristic scale $L^{\mbox{\tiny{(0)}}}$ (the size of the initial slip pulse). In the other limit, where the characteristic spatial scale of variation of $\tau_{\rm b}(x)$ is smaller than $L^{\mbox{\tiny{(0)}}}$, we expect the pulse to smooth out the spatial variability of the prestress (see Sect.~\ref{sm:small-lambda} in the SM).
%%%%%%%%%%%%%%%%%%%%%%%%%%%%%%%%%%%%%%%
\begin{figure*}[ht!]
\center
\includegraphics[width=0.99\textwidth]{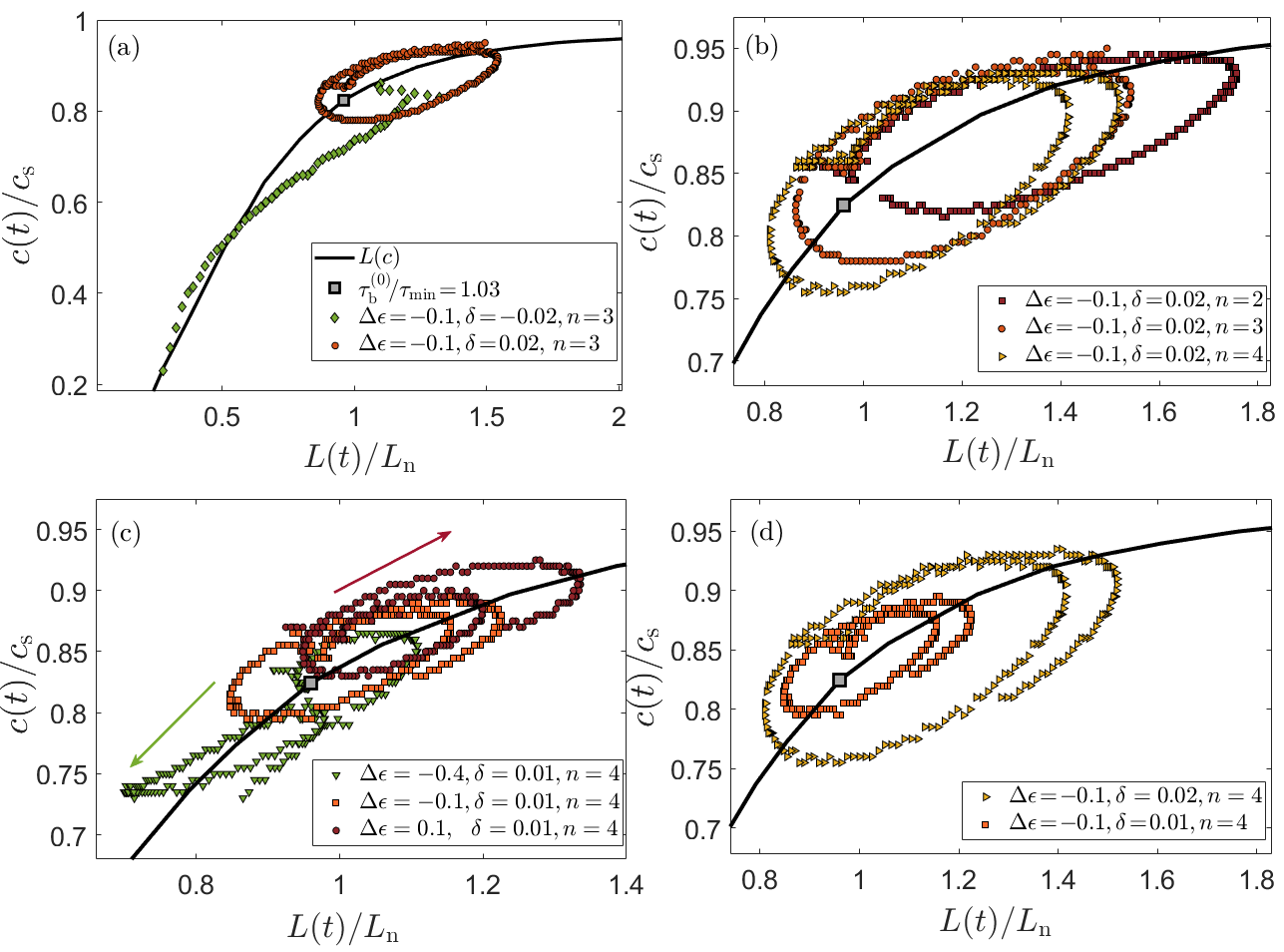}\
\caption{Unsteady pulse dynamics under periodic prestress $\tau_{\rm b}(x)$ of Eq.~\eqref{eq:sin}, with fault average $\langle\tau_{\rm b}(x)\rangle_x/\tau_{\rm min}\!=\!1.03$. In all panels, the square represents the $\tau^{\mbox{\tiny{(0)}}}_{\rm b}/\tau_{\rm min}\!=\!1.03$ steady-state pulse and the black line represents the special $L(c)$ line in the $L\!-\!c$ plane. (a) Pulse dynamics for $\Delta \epsilon\!=\!-0.1$ and $n\!=\!3$, with $\delta\!=\!0.02$ (orange circles) and $\delta\!=\!-0.02$ (green diamonds). See legend and text for discussion. (b) Pulse dynamics for $\Delta\epsilon\!=\!-0.1$ and $\delta\!=\!0.02$, with different prestress wavelengths $\lambda\!=\!W/n$ (varied through $n$). See legend and text for discussion. (c) Pulse dynamics for $\delta\!=\!0.01$ and $n\!=\!4$, with different initial instability indices $\Delta\epsilon$ (see legend). The green and brown arrows indicate the overall direction of the pulse trajectory in the $L\!-\!c$ plane. See text for discussion. (d) Pulse dynamics for $\Delta\epsilon\!=\!-0.1$ and $n\!=\!4$, with different positive amplitudes $\delta\!>\!0$ (see legend and text for discussion).}
\label{fig:fig2}
\end{figure*}
%%%%%%%%%%%%%%%%%%%%%%%%%%%%%%%%%%%%%%%

\section{P\lowercase{ulse dynamics under periodic prestress}}
\label{sec:periodic}

We first consider the periodic prestress distributions in Eq.~\eqref{eq:sin}. We set $\langle\tau_{\rm b}(x)\rangle_x/\tau_{\rm min}\!=\!1.03$ and study pulse dynamics in a system with $W\!=\!480$ m, for various $\Delta\epsilon$, $\delta$ and $n$. In Fig.~\ref{fig:fig2}a, we present the pulse dynamics in the $L\!-\!c$ plane for $\Delta\epsilon\!=\!-0.1$, i.e., a pulse that decays under a uniform prestress. For $\delta\!=\!0.02$ and $n\!=\!3$, it is observed that the pulse sustains, instead of decaying, and forms a limit cycle in the $L\!-\!c$ plane (orange circles). The limit cycle, corresponding to a pulse whose $L(t)$ and $c(t)$ vary nearly periodically as it propagates (see Fig.~\ref{fig:v_snap_and_vmax} in the SM), remains close to the special $L(c)$ line (solid black line) and encircles the corresponding uniform prestress fixed point (square). In this case, the periodic prestress transforms a decaying pulse (obtained under a uniform prestress corresponding to the fault average of the periodic one) into a long-lived, sustained pulse. Changing the overall sign of the periodic prestress (i.e., setting $\delta\!=\!-0.02$), leads to a decaying pulse, which closely follows the special $L(c)$ line (green diamonds). These results indicate that the gradient of $\tau_{\rm b}(x)$ near $x\!=\!0$, in particular its sign, plays an important role in determining the fate of the pulse (i.e., whether it decays or sustains and follows a limit cycle in the $L\!-\!c$ plane).

In Fig.~\ref{fig:fig2}b, we set $\Delta\epsilon\!=\!-0.1$ and $\delta\!=\!0.02$, and vary the wavelength of the periodic prestress (by setting $n\!=\!2,3,4$, see legend). It is observed that sustained pulses, forming quasi-limit cycles in the $L\!-\!c$ plane close to the special $L(c)$ line, emerge. The quasi-limit cycles are pushed to larger $L$ and $c$ values with increasing wavelength (decreasing $n$). We use the term ``quasi-limit cycle'' as in some cases the limit cycle is not exact and since we did not always follow the dynamics over very long times (and system sizes $W$), due to computational constraints; consequently, we cannot always determine whether the limit cycle is exact or approximate (e.g., drifts with increasing time, see Fig.~\ref{fig:v_snap_and_vmax} in the SM).

In Fig.~\ref{fig:fig2}c, we fix the periodic prestress parameters $\delta\!=\!0.01$ and $n\!=\!4$, and vary the instability index $\Delta\epsilon$ of the initial pulse. For $\Delta\epsilon\!=\!0.1$ (brown cycles), a quasi-limit cycle close to the special $L(c)$ line emerges (the cycle in the $L\!-\!c$ plane seems to be drifting upwards with time). Upon decreasing $\Delta\epsilon$, setting it to $\Delta\epsilon\!=\!-0.1$ (orange squares), a quasi-limit cycle also appears to emerge (recall that $\delta\!>\!0$). Upon further decreasing $\Delta\epsilon$, setting it to $\Delta\epsilon\!=\!-0.4$ (green triangles), the pulse initially attempts to form a limit cycle in the $L\!-\!c$ plane, but appears to be drifting to smaller $L$ and $c$ values, and to decay. Finally, in Fig.~\ref{fig:fig2}d, we fix $\Delta\epsilon\!=\!-0.1$ and $n\!=\!4$, and vary the amplitude $\delta\!>\!0$ of the periodic prestress (see legend). It is observed that the emerging quasi-limit cycle broadens with increasing $\delta\!>\!0$. Overall, the results presented in Fig.~\ref{fig:fig2} indicate that a periodic prestress distribution can reverse the fate of the original pulse (under uniform prestress), leading to either decaying pulses or to quasi-limit cycles in the $L\!-\!c$ plane. In both cases, for the parameters considered, the unsteady pulse dynamics remain rather close to the special $L(c)$ line, which corresponds to steady-state pulse solutions parameterized by a uniform pretress $\tau^{\mbox{\tiny{(0)}}}_{\rm b}$~\cite{pomyalov2023dynamics}.
%%%%%%%%%%%%%%%%%%%%%%%%%%%%%%%%%%%%%%%
\begin{figure}[ht!]
\center
\includegraphics[width=0.49\textwidth]%
{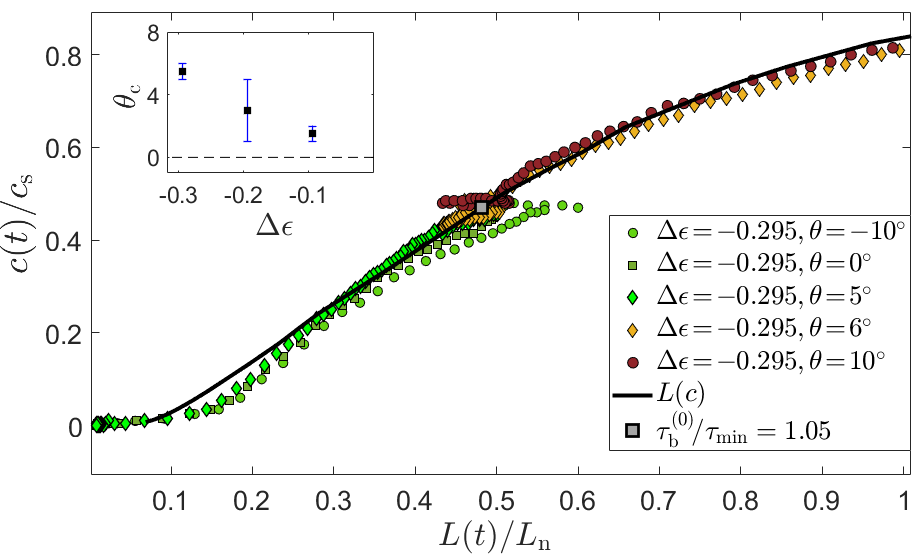}
\caption{Unsteady pulse dynamics under constant-gradient prestress $\tau_{\rm b}(x)$ of Eq.~\eqref{eq:tilt}, with fault average $\langle\tau_{\rm b}(x)\rangle_x/\tau_{\rm min}\!=\!1.05$. Results for $\Delta\epsilon\!=\!-0.295$ and various tilt angles in the range $-10^\circ\!\le\!\theta\!\le\!10^\circ$ are presented, see legend and text for discussion. (inset) The threshold tilt angle $\theta_{\rm c}\!>\!0$ (with error bars), required to induce a transition from decaying to growing/expanding pulses, as a function of $\Delta\epsilon\!<\!0$.}
\label{fig:fig3}
\end{figure}
%%%%%%%%%%%%%%%%%%%%%%%%%%%%%%%%%%%%%%%
%%%%%%%%%%%%%%%%%%%%%%%%%%%%%%%%%%%%%%
\begin{figure*}[ht!]
\center
\includegraphics[width=0.99\textwidth]{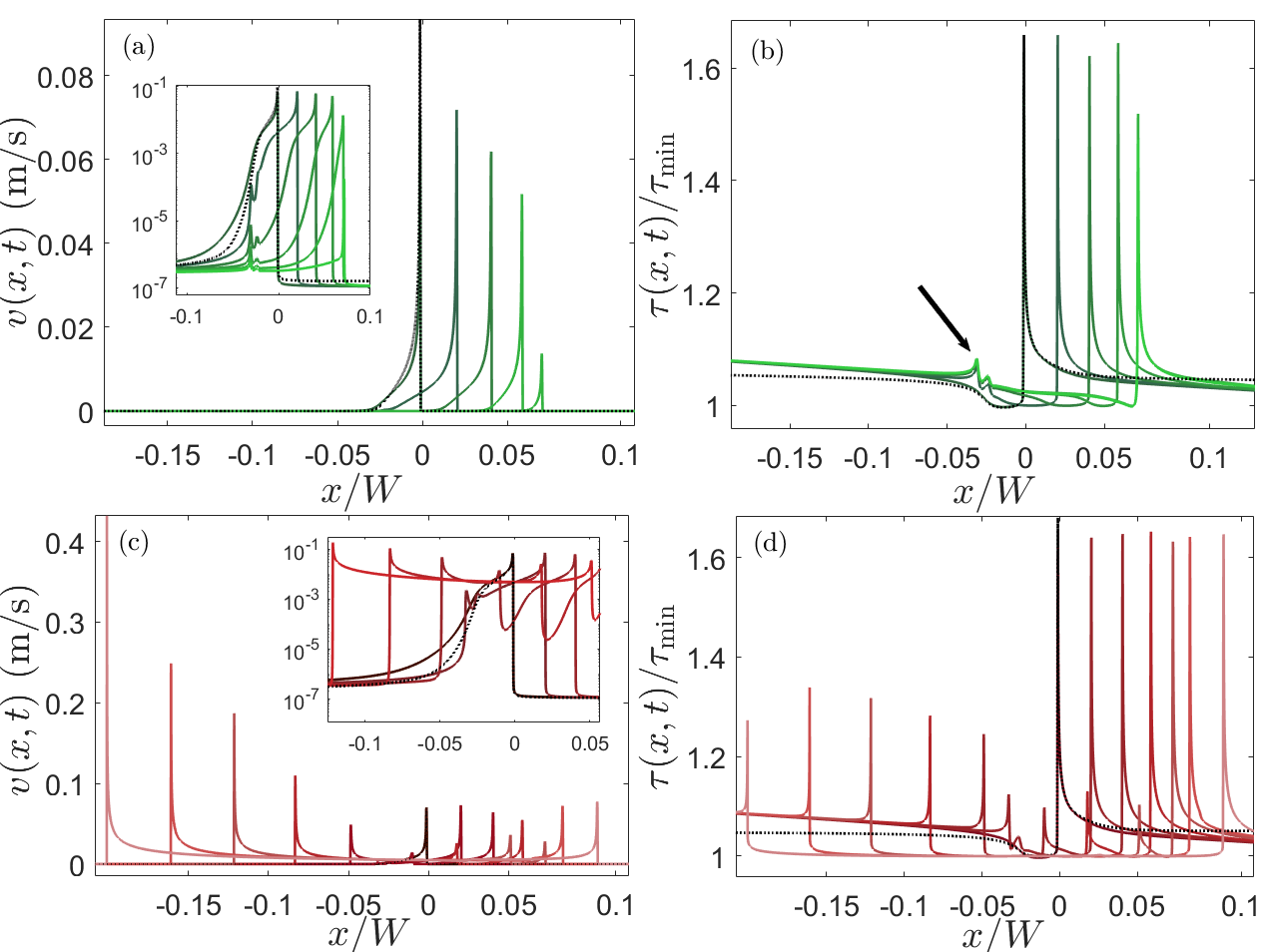}
\caption{A series of equal-time snapshots of the slip velocity field $v(x,t)$ (left column) and shear stress field $\tau(x,t)$ (right column), obtained for $\langle\tau_{\rm b}(x)\rangle_x/\tau_{\rm min}\!=\!1.05$ and an initial pulse with $\Delta\epsilon\!=\!0.005$ (shown in a dotted black line in each panel), under a constant-gradient prestress with $\theta\!=\!-9^\circ$ (top row) and $\theta\!=\!-10^\circ$ (bottom row). Time evolves from darker to lighter colors, also revealing the direction of rupture propagation. The insets in panels (a) and (c) present a spatial zoom-in of the main panels, and employ a logarithmic $y$-axis for the slip velocity field $v(x,t)$. Note also that pulse arrest leaves behind it a heterogeneous stress distribution (brightest, rightmost snapshot in panel (b)). The presented results are extensively discussed in the text. See also Fig.~\ref{fig:fig5}, which presents the results of panel (c) in a space-time plot.}
\label{fig:fig4}
\end{figure*}
%%%%%%%%%%%%%%%%%%%%%%%%%%%%%%%%%%%%%%%

\section{P\lowercase{ulse dynamics under constant-gradient prestress}}
\label{sec:constant_gradient}

We next consider the constant-gradient prestress distributions in Eq.~\eqref{eq:tilt}. We set $\langle\tau_{\rm b}(x)\rangle_x/\tau_{\rm min}\!=\!1.05$ and study pulse dynamics in a system with $W\!=\!480$ m, for various instability indices $\Delta\epsilon$ and tilt angles $\theta$. In Fig.~\ref{fig:fig3}, we present the pulse dynamics in the $L\!-\!c$ plane for $\Delta\epsilon\!=\!-0.295$, i.e., a pulse that decays under a uniform prestress (corresponding to $\theta\!=\!0^\circ$ in the figure, see green squares), and various values of $\theta$. Upon imposing a negative prestress gradient, corresponding to $\theta\!=\!-10^\circ$ (green circles), the pulse decays with a small down shift compared to the reference $\theta\!=\!0^\circ$ results, yet it still follows rather closely the special $L(c)$ line (solid black line). Upon imposing a positive prestress gradient of $\theta\!=\!5^\circ$ (green diamonds), the pulse decays with a small up shift compared to the reference $\theta\!=\!0^\circ$ results.

The dynamics qualitatively change upon setting $\theta\!=\!6^\circ$ (yellow diamonds), where the pulse changes its character and becomes a growing/expanding pulse, which closely follows the special $L(c)$ line. A similar growing/expanding pulse is observed for $\theta\!=\!10^\circ$ (brown circles), featuring a small up shift compared to the $\theta\!=\!6^\circ$ results. These results indicate that for a given instability index $\Delta\epsilon\!<\!0$, there exists a threshold tilt angle $\theta_{\rm c}\!>\!0$ that can reverse the nature/fate of the pulse, from decaying to growing. For $\Delta\epsilon\!=\!-0.295$, $\theta_{\rm c}$ is between $5^\circ$ and $6^\circ$. In the inset of Fig.~\ref{fig:fig3}, we plot $\theta_{\rm c}$ against $\Delta\epsilon\!<\!0$, demonstrating that there is an interplay between the instability index of the initial pulse and the threshold tilt angle $\theta_{\rm c}\!>\!0$ needed to induce a transition from decaying to growing/expanding pulses.

Our analysis above focused on initial self-healing slip pulses that remain pulses under various spatially-varying prestress conditions. Yet, as noted in Sect.~\ref{sec:intro}, earthquakes can take the form of either pulse-like or crack-like rupture, with significant implications for the duration of slip, earthquake energy budget, rupture scaling laws and seismic radiation. The dynamic selection of these rupture modes/styles and the spontaneous transition between them are important topics under active investigation, e.g., see~\cite{Perrin1995,lu2007pulse,Gabriel2012,Brener2018,brantut2019stability,heimisson2020crack}. It is generally established~\cite{Perrin1995,lu2007pulse,Gabriel2012,Brener2018,brantut2019stability} that crack-like rupture is more prevalent under larger average prestress levels. Consequently, it would be interesting to see whether for $\langle\tau_{\rm b}(x)\rangle_x/\tau_{\rm min}\!=\!1.05$ (a higher average prestress compared to the one considered in Sect.~\ref{sec:periodic}), a spatially-varying prestress distribution can spontaneously induce a pulse-to-crack rupture transition.

To address this question, we present in Fig.~\ref{fig:fig4} (top row) a series of equal-time snapshots of $v(x,t)$ (Fig.~\ref{fig:fig4}a) and $\tau(x,t)$ (Fig.~\ref{fig:fig4}b) obtained for $\langle\tau_{\rm b}(x)\rangle_x/\tau_{\rm min}\!=\!1.05$ and $\Delta\epsilon\!=\!0.005$, under a constant-gradient prestress with $\theta\!=\!-9^\circ$. It is observed that the nonuniform prestress transforms a growing pulse ($\Delta\epsilon\!>\!0$) into a decaying pulse, which propagates a certain distance from left to right (in the positive $x$ direction), until it arrests/dies off. Note that pulse arrest leaves behind it a heterogeneous stress distribution (brightest, rightmost snapshot in panel (b)). Interestingly, it is observed that near the healing tail of the initial pulse, around $x/W\!\simeq\!-0.03$ (also marked by an arrow), there seems to be a spontaneous attempt to nucleate secondary rupture. The latter is evident in $\tau(x,t)$ (Fig.~\ref{fig:fig4}b), but requires a logarithmic scale to be discerned in $v(x,t)$, as done in the inset of Fig.~\ref{fig:fig4}a.

In Fig.~\ref{fig:fig4} (bottom row), we present the spatiotemporal dynamics for a slightly stronger negative gradient, corresponding to $\theta\!=\!-10^\circ$ (everything else being the same as in the top row). It is observed that this nonuniform prestress leads to a qualitative change in the dynamics. First, the attempted nucleation near the healing tail of the initial pulse, around $x/W\!\simeq\!-0.03$, becomes a successful nucleation event, giving rise to an asymmetric bilateral crack-like rupture. The rupture edge that propagates from right to left (in the negative $x$ direction) corresponds to a back-propagating rupture. The rupture edge propagating in the forward (positive $x$) direction, catches up with the initial pulse that also propagates in the forward direction and merges with it. Second, upon merging of the two forward-propagating rupture fronts, a persistent asymmetric bilateral crack-like rupture emerges. Overall, it is observed that a spatially-varying prestress can induce a pulse-to-crack transition.
%%%%%%%%%%%%%%%%%%%%%%%%%%%%%%%%%%%%%%
\begin{figure}[ht!]
\center
\includegraphics[width=0.48\textwidth]{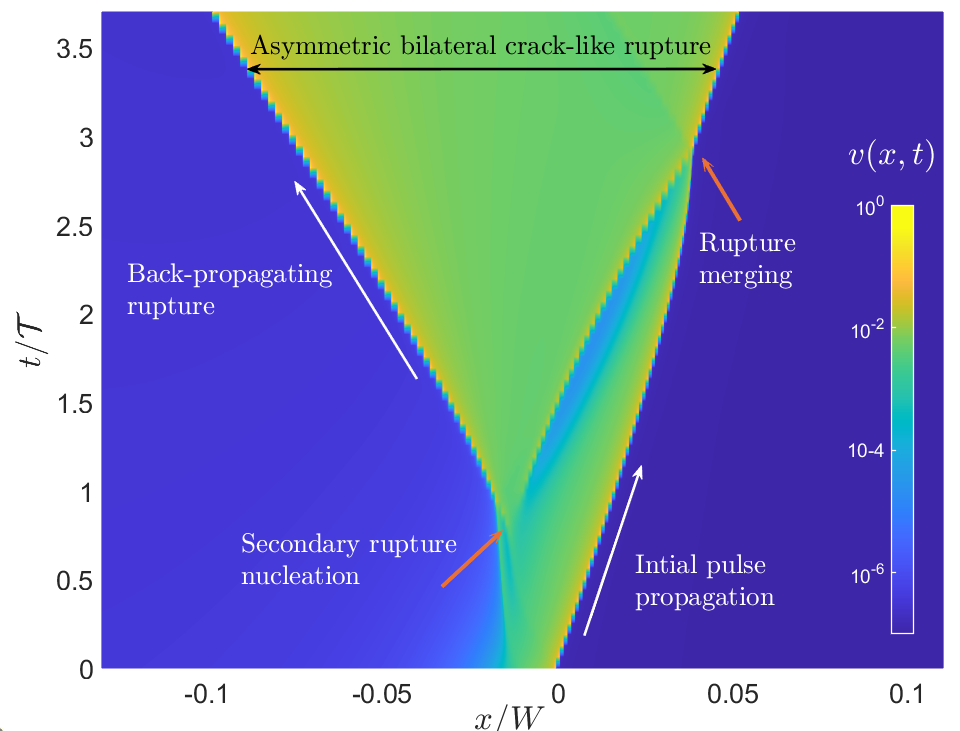}
\caption{Results of the same computation as in Fig.~\ref{fig:fig4}c-d, presented here in a space-time $x\!-\!t$ plot. The slip velocity field $v(x,t)$ (in units of m/s, as in Fig.~\ref{fig:fig4}c) is plotted on a logarithmic scale, see color legend on the right. $x$ is normalized by the fault size $W$ and $t$ by ${\cal T}\!\equiv\!L^{\mbox{\tiny{(0)}}}/c^{\mbox{\tiny{(0)}}}$. All dynamical events and various rupture propagation modes are marked, see text for additional discussion.}
\vspace{-0.5cm}
\label{fig:fig5}
\end{figure}
%%%%%%%%%%%%%%%%%%%%%%%%%%%%%%%%%%%%%%%

The crack-like nature of the two persistent, oppositely propagating rupture edges is evident in the inset of Fig.~\ref{fig:fig4}c (note the logarithmic $y$-axis), where a persistent finite slip velocity is left in the central part, while the two oppositely propagating rupture edges move further away. Similarly, the crack-like nature of the emerging rupture is also manifested in the residual stress observed in Fig.~\ref{fig:fig4}d. These results, which are presented again in a space-time plot in Fig.~\ref{fig:fig5}, demonstrate that rich spatiotemporal rupture dynamics may emerge from rather simple spatially-varying prestress distributions, revealing phenomena such as back-propagating rupture and pulse-to-crack transitions, which are of interest in various geophysical contexts, e.g., in relation to earthquake source models and seismological inversion analyses~\cite{idini2020fault,ding2024back}. Finally, we note --- as highlighted in~\cite{pomyalov2023dynamics} --- that once pulse-like rupture transforms into a crack-like rupture, the dynamics of latter can no longer be meaningfully represented in the $L\!-\!c$ plane and an intrinsic relation to the special $L(c)$ line --- which is essentially an equation of motion for self-healing slip pulses under uniform prestress --- does not exist anymore.

\section{S\lowercase{ummary and concluding remarks}}
\label{sec:summary}

In this paper, we studied the interaction of well-defined initial self-healing slip pulses with two classes of spatially-varying prestress distributions $\tau_{\rm b}(x)$, one corresponding to a periodic variation and one to a constant gradient, see Eqs.~\eqref{eq:sin}-\eqref{eq:tilt}. Our findings can be summarized along two major axes. The first concerns the richness (or to some extent the complexity) of rupture dynamics emerging from initial slip pulses interacting with a spatially-varying prestress (compared to their uniform prestress counterparts). The second concerns the theoretical description of unsteady slip pulse dynamics under spatially-varying prestress, particularly in relation to the recently established reduced-dimensionality description of pulse dynamics~\cite{pomyalov2023dynamics}.

We have demonstrated that simple nonuniform prestress distributions can give rise to rather rich rupture dynamics. In particular, we showed that a nonuniform prestress can reverse the nature/fate of a slip pulse, from decaying/shrinking to growing/expanding and vice versa, for a fixed fault-averaged prestress, see Figs.~\ref{fig:fig2}-\ref{fig:fig3}. We also showed that pulses under periodic prestress can settle into propagating states in which their properties, e.g., pulse size $L(t)$, peak slip velocity $v_{\rm p}(t)$ and propagation velocity $c(t)$, vary periodically or quasi-periodically as they propagate. Finally, we demonstrated that prestress spatial gradients can lead to the spontaneous nucleation of secondary rupture near the healing tail of a pulse, to back-propagating rupture and to pulse-to-crack transitions, see Figs.~\ref{fig:fig4}-\ref{fig:fig5}.

The above mentioned rupture phenomena are geophysically documented phenomena of importance, which continue to pose challenges to our understanding. For example, it is known that while most seismic studies typically only estimate the average rupture velocity, seismic radiation is predominantly affected by the spatiotemporal variation of the rupture velocity~\cite{Madariaga1977}. Our finding, e.g., in relation to the emergence of quasi-periodically propagating slip pulses, can contribute in this context. Moreover, as noted in~\cite{elbanna2011pulselike}, slip pulse propagation and arrest give rise to heterogeneous after-event stresses (cf., top row of Fig.~\ref{fig:fig4}) and hence may offer a mechanism for sustaining prestress heterogeneity over many earthquake cycles.

On the more theoretical side, our results demonstrate that the reduced-dimensionality description of unsteady pulses, originally established under uniform prestress conditions~\cite{pomyalov2023dynamics}, is still relevant under some nonuniform prestress conditions. Specifically, it has been shown that such pulse dynamics presented in a plane defined by a pair of pulse properties --- e.g., its size $L(t)$ and propagation velocity $c(t)$ ---, remain close to a special, previously derived line $L(c)$. For example, periodically (or quasi-periodically) propagating slip pulses under nonuniform prestress correspond to limit cycles (or quasi-limit cycles) in the $L\!-\!c$ plane that remain rather close to the special $L(c)$ line at all times, see Figs.~\ref{fig:fig2}-\ref{fig:fig3} and SM.

Consequently, while the special $L(c)$ line is not a strict equation of motion for unsteady pulses under nonuniform prestress, in many cases it does provide a good approximation for the dynamics, in the sense that the knowledge of one pulse property, e.g., $c(t)$, can be used to infer another property, e.g., $L(t)$. Finally, we also demonstrated that similarly to the uniform prestress case~\cite{pomyalov2023dynamics}, the emerging theoretical picture remains valid upon using other pairs of pulse properties, e.g., $L(t)$ and the peak slip velocity $v_{\rm p}(t)$ (see SM).\\

{\em Acknowledgements}. We acknowledge support by the Israel Science Foundation (ISF grant no.~1085/20), the Minerva Foundation (with funding from the Federal German Ministry for Education and Research), the Ben May Center for Chemical Theory and Computation, and the Harold Perlman Family. We are grateful to Prof.~Eric Dunham for posing to us the basic question, to Dr.~Fabian Barras for his help with setting up the initial conditions in cRacklet and to Dr.~Efim Brener for useful discussions.

%\newpage

\clearpage

\onecolumngrid
%\vspace{1cm}
\begin{center}
              \textbf{\Large Supplementary material}
\end{center}

%%%%%%%%%%%%%%%%%%%%%%%%%%%%%%%%%%%%%%%%%%%%%%%%%%%%%%%%%%%%%%%%%%%%%%%%%%%%%%%%%
%%%%%%%%%%%%%%%%%%%%%% these lines of code handle the concatenation %%%%%%%%%%%%%
%%%%%%%%%%%%%%%%%%%%%%%%%%%%%%%%%%%%%%%%%%%%%%%%%%%%%%%%%%%%%%%%%%%%%%%%%%%%%%%%%
\setcounter{equation}{0}
\setcounter{figure}{0}
\setcounter{section}{0}
\setcounter{subsection}{0}
\setcounter{table}{0}
\setcounter{page}{1}
\makeatletter
\renewcommand{\theequation}{S\arabic{equation}}
\renewcommand{\thefigure}{S\arabic{figure}}
\renewcommand{\thesection}{S-\Roman{section}}
\renewcommand*{\thepage}{S\arabic{page}}
\renewcommand{\thetable}{S-\arabic{table}}
%\renewcommand{\bibnumfmt}[1]{[S#1]}
%\renewcommand{\bibnumfmt}[1]{[S#1]}
%\renewcommand{\citenumfont}[1]{S#1}
%%%%%%%%%%%%%%%%%%%%%%%%%%%%%%%%%%%%%%%%%%%%%%%%%%%%%%%%%%%%%%%%%%%%%%%%%%%%%%%%%
%%%%%%%%%%%%%%%%%%%%%% these lines of code handle the concatenation %%%%%%%%%%%%%
%%%%%%%%%%%%%%%%%%%%%%%%%%%%%%%%%%%%%%%%%%%%%%%%%%%%%%%%%%%%%%%%%%%%%%%%%%%%%%%%%

The goal of this document is to provide some technical details regarding the results presented in the manuscript and to offer some additional supporting data.

\vspace{-0.42cm}
\section{T\lowercase{he interfacial constitutive relation (friction law)}}
\label{sec:friction_law}

The left-hand-side of Eq.~\eqref{eq:BIM} corresponds to the frictional strength $\tau[v,\phi]$, where $v(x,t)$ is the slip velocity field and $\phi(x,t)$ is the internal state field. Exactly as in~\citep{Brener2018,pomyalov2023self,pomyalov2023dynamics}, we have
\begin{eqnarray}
\label{eq:f}
\tau(v,\phi)/\sigma = \left[1+b\log\left(1+\phi/\phi^{\star}\right)\right] \left[f_0/\sqrt{1+\left(v^{\star}\!/v\right)^2}+\alpha\,\log\left(1+{\abs  v}/v^{\star}\right)\right]\ ,
\end{eqnarray}
where the frictional parameters appearing in the expression are given in the table in Fig.~\ref{fig:fig1sm}(left). This table also reports on the linear elastic bulk parameters of Eq.~\eqref{eq:BIM} (the spatiotemporal convolutional integral $s(x,t)$ therein features in addition to $\mu$ and $c_{\rm s}$, which appear in the radiation-damping term, also Poisson's ratio $\nu$) and the value of the applied normal stress $\sigma$. As explained in the manuscript, the internal state field $\phi(x,t)$ evolves according the ``aging law'', taking the form $\partial_t\phi(x,t)\!=\!1\!-\!v(x,t)\,\phi(x,t)\sqrt{1+v^*/v(x,t)}/D$, which includes the characteristic slip distance $D$, whose value is also given in the table in Fig.~\ref{fig:fig1sm}(left). Note that the regularization factor $\sqrt{1+v^*/v(x,t)}$ was not included in the ``aging law'' expression in the manuscript, as it is relevant only at extremely small slip velocities (note the value of $v^*$ in the table in Fig.~\ref{fig:fig1sm}). Under spatially homogeneous steady sliding, we have $\partial_t\phi\!=\!0$, which allows to express $\phi_{\rm ss}$ (the steady sliding value of the internal state field) in terms of $v_{\rm ss}$ (the steady sliding value of the slip velocity field). Once done, one can obtain the steady sliding friction curve $\tau(v_{\rm ss})$, plotted in Fig.~\ref{fig:fig1sm}(right). $\tau(v_{\rm ss})$ follows a generic N shape, featuring a local minimum at $(v_{\rm min}, \tau_{\rm min})$~\cite{Baumberger2006Solid,Bar-Sinai2014} (marked on the figure). The existence of a minimum, though, does not play a crucial role in the problem we solved; yet, the characteristic slip velocity $v_{\rm min}\!=\!0.005971$ m/s and characteristic frictional strength $\tau_{\rm min}\!=\!0.3406679$ MPa can be used to nondimensionalize various quantities, as done in the manuscript. We also note that using the previously-calculated effective fracture energy $G_{\rm c}\!=\!0.65$~J/m$^2$ of slip pulses~\cite{pomyalov2023self} for our frictional parameters, one can construct a characteristic pulse length as $L_{\rm n}\!=\!(c_{\rm s}/v_{\rm min})^2 (G_{\rm c}/\mu)\!=\!15.2$~m. The latter is used to normalize the size of slip pulses $L(t)$ in the manuscript.

\section{P\lowercase{erturbed steady-state pulses as initial conditions}}
\label{sub:initial_conditions}

Our main goal in the manuscript is to systematically and controllably study the interaction of self-healing slip pulses with spatially-varying prestress distributions. To that aim, we need well-defined and controlled initial conditions that correspond to slip pulses. Here, we explain how this is achieved, providing additional details on top of the discussion offered in the manuscript.

\subsection{Steady-state self-healing pulses in finite periodic domains}
\label{subsec:finite_periodic}

Similarly to~\cite{pomyalov2023dynamics}, we employ the steady-state slip pulse solutions recently obtained under a uniform prestress in~\cite{pomyalov2023self}. The latter solutions take the form $v_{_{\rm ss}}\!(\xi)$ and $\phi_{_{\rm ss}}\!(\xi)$, with $\xi\!=\!x-c^{\mbox{\tiny{(0)}}}t$ being a co-moving space-time coordinate $-\infty\!<\!\xi\!<\!\infty$ and $c^{\mbox{\tiny{(0)}}}$ is the steady pulse propagation velocity, see Fig.~\ref{fig:fig1}a, where the finite size of the steady-state pulse, $L^{\mbox{\tiny{(0)}}}$, is also marked. Specifically, for a given spatially-varying prestress distribution $\tau_{\rm b}(x)$, we use a steady-state pulse solution (or a slightly perturbed version of it quantified by the instability index $\Delta\epsilon$, see Sect.~\ref{subsec:perturbation}) corresponding to a uniform prestress of magnitude $\langle\tau_{\rm b}(x)\rangle_x$, i.e., the spatial fault average of $\tau_{\rm b}(x)$, at $t\!=\!0$. The latter is used as an initial condition for the $t\!>\!0$ spatiotemporal dynamics described by the boundary integral equation (Eq.~\eqref{eq:BIM}), which is solved using the open-source library $cRacklet$~\cite{roch2022cracklet}. $cRacklet$ is a boundary element library based on a spectral formulation of linear elastodynamics~\cite{Breitenfeld1998}, and as such is applicable in finite domains of size $W$ in the fault direction $x$ (the domain is infinite in the perpendicular $y$ direction, the fault is located at $y\!=\!0$), where periodic boundary conditions are employed.

The periodic domain of size $W$ is clearly in conflict with steady-state pulses that propagated steadily at a velocity $c^{\mbox{\tiny{(0)}}}$ indefinitely in the past, $-\infty\!<\!t\!<\!0$. That is, generating slip history (past propagation) that extends beyond the domain boundary implies that the past spuriously affects future dynamics. The way to resolve this conflict has been discussed in detail in~\cite{pomyalov2023dynamics}, see in particular section S-3A of the Supplemental Materials file therein. In a nutshell, we require that the past propagation distance of the steady-state pulse, denoted by $w$, is much smaller than the spatial periodicity domain $W$ and much larger than the steady-state pulse size $L^{\mbox{\tiny{(0)}}}$, i.e., $L^{\mbox{\tiny{(0)}}}\!\ll\!w\!\ll\!W$ (see Fig.~S1 in~\cite{pomyalov2023dynamics}). Effectively, this means that instead of indefinite past propagation (i.e., for $-\infty\!<\!t\!<\!0$), we account for finite time past propagation over $-t_0\!<\!t\!<\!0$, where $t_0\!>\!0$ is determined by the spatial position in which the past propagation is initiated. Once a given slip history is generated, it is used to calculate the spatiotemporal convolutional integral and the resulting field $s(x,t\!=\!0)$ is used in Eq.~\eqref{eq:BIM} as an initial condition. The latter can be supplemented with a perturbation in the initial internal state field, quantified by the instability index $\Delta\epsilon$, as will be discussed in Sect.~\ref{subsec:perturbation}.
%%%%%%%%%%%%%%%%%%%%%%%%%%%%%%%%%%%%%%%%%%%%%%
%minipage environment including a table and a figure with a common caption
%%%%%%%%%%%%%%%%%%%%%%%%%%%%%%%%%%%%%%%%%%%%%%%%%%
\begin{table*}
\begin{minipage}{0.4\textwidth}
\centering
\begin{tabular}{|c|c|c|}
\hline
$\sigma$ & $1$  & MPa\\
	\hline
$\mu$            & $9$&	GPa\\
\hline
$\nu$            & $0.33$&--\\
\hline
$c_{\rm s} $      &$2739$  &m/s\\
\hline
 $D$              &$5\times 10^{-7}$ & m\\
 \hline
  $v^*$           &$10^{-7}$  &m/s\\
  \hline
$\phi^*$         &$3.3\times10^{-4}$ & s\\
\hline
$\alpha$         & $0.005$	&--\\
\hline
$b$              &$0.075$  &--\\
\hline
  $f_0$          &$0.28$ & --\\
	\hline
\end{tabular} \label{t:1}
%\caption{The normal stress $\sigma$, linear elastic bulk parameters and frictional parameters (cf.~Eqs.~\eqref{eq:BIM}-\eqref{eq:f}) used in the computer simulations.}
\end{minipage}
\hfill
\begin{minipage}{0.57\textwidth}
\includegraphics[width=\textwidth]{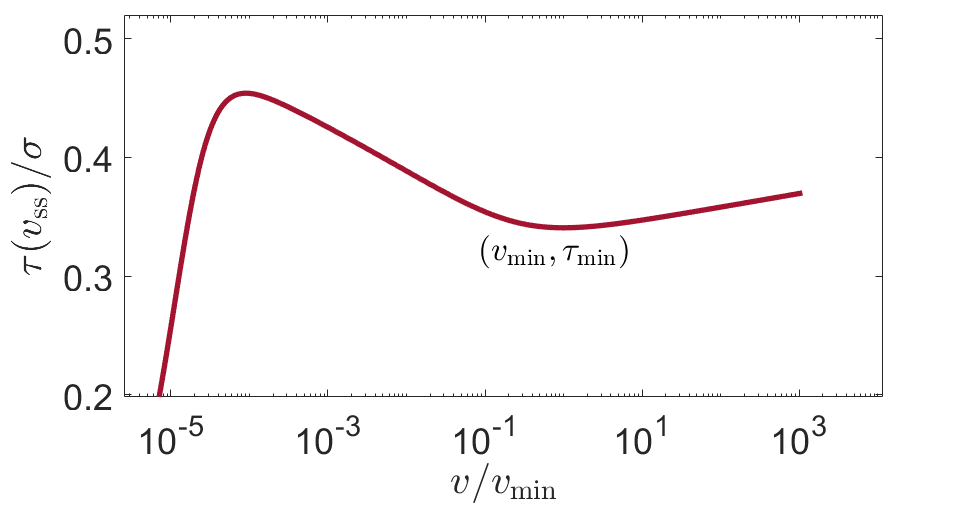}
\end{minipage}
\captionof{figure}{(left) The applied normal stress $\sigma$, the linear elastic bulk parameters (entering the right-hand-side of Eq.~\eqref{eq:BIM}) and the frictional parameters (cf.~Eq.~\eqref{eq:f} and the $\phi$ evolution equation) used in this work. (right) The steady frictional resistance $\tau(v_{\rm ss})$, see text. The local minimum of the curve $(v_{\rm min}, \tau_{\rm min})$ is marked.}
\label{fig:fig1sm}
\end{table*}

\begin{figure*}[ht!]
\center
\includegraphics[width=\textwidth]%
%{StaticStressCorrection_1.05_epsilon_m0.3.png}
{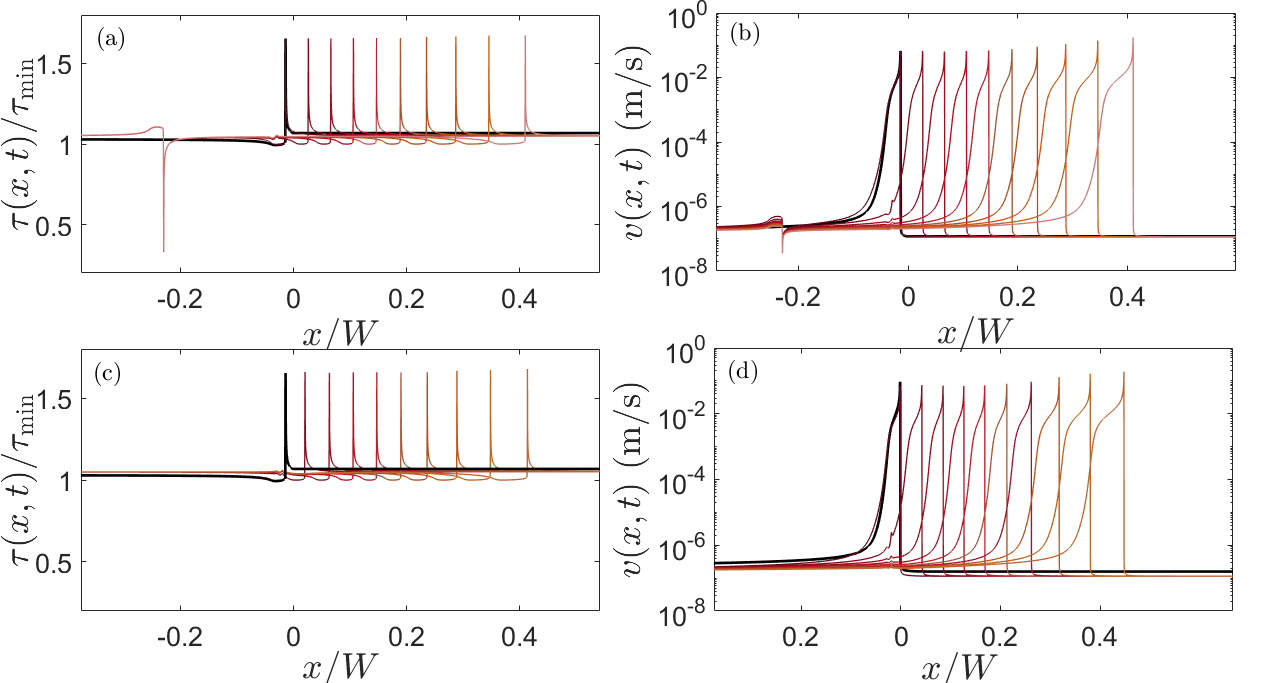}
\caption{(a) The spatiotemporal evolution of the shear stress field $\tau(x,t)$, during pulse propagation corresponding to one of the solutions discussed in~\cite{pomyalov2023dynamics} (with a uniform prestress $\tau^{\mbox{\tiny{(0)}}}_{\rm b}/\tau_{\rm min}\!=\!1.05$ and $\Delta\epsilon\!=\!0.005$), in the presence of the frozen-in, artificial stress concentration (appear at $x/W\!\simeq\!-0.25$), see text for extensive discussion. The desired initial slip pulse, which does not feature the artificial stress concentration, appears in the thick black curve. The different snapshots illustrate the propagation of the pulse from left (dark colors) to right (light colors). (b) The corresponding spatiotemporal evolution of the slip velocity field $v(x,t)$ (note the semi-log scale). Panels (c) and (d) present the corresponding spatiotemporal dynamics once the frozen-in, artificial stress concentration is removed (see text).}
\label{fig:static_stress_correciton}
\end{figure*}

The above-described procedure for mimicking an initial condition corresponding to a propagating steady-state pulse in a finite, periodic system has one implication that has not been discussed in~\cite{pomyalov2023dynamics}. Specifically, the fact that the past indefinite propagation over $-\infty\!<\!t\!<\!0$ is replaced by a finite past propagation over $-t_0\!<\!t\!<\!0$ gives rise to a frozen-in and localized artificial stress concentration due to the unaccounted for slip accumulated over $-\infty\!<\!t\!<\!-t_0$. To demonstrate the point, we present in Fig.~\ref{fig:static_stress_correciton}a the spatiotemporal evolution of the shear stress field $\tau(x,t)$, corresponding to one of the solutions discussed in~\cite{pomyalov2023dynamics} with a uniform prestress. The desired slip pulse initial stress distribution appears in the thick black curve. It is observed that the latter is indeed realized, except for a localized deviation near $x/W\!\simeq\!-0.25$. The latter persists as time progresses, upon which the pulse propagates to the right, further away from the artificial stress concentration. The corresponding dynamics of the slip velocity field $v(x,t)$ are presented in Fig.~\ref{fig:static_stress_correciton}b. In~\cite{pomyalov2023dynamics}, we made sure that the frozen-in, artificial stress concentration was located sufficiently away from the central part of the slip pulse such that the dynamics were not affected. The artifact can be readily removed by applying a static slip that exactly relaxes this artificial stress concentration at $t\!=\!0$, as done in Fig.~\ref{fig:static_stress_correciton}c-d. Comparing the top and bottom rows of Fig.~\ref{fig:static_stress_correciton}, we indeed observe that the frozen-in, artificial stress concentration does not affect the pulse dynamics. Nevertheless, in the present work, we eliminate it in generating the initial condition, at $t\!=\!0$, under uniform prestress. The spatially-varying prestress distribution $\tau_{\rm b}(x)$ is introduced at $t\!=\!0^+$, as explained in the manuscript.
%%%%%%%%%%%%%%%%%%%%%%%%%%%%%%%%%%%%%%%
%%%%%%%%%%%%%%%%%%%%%%%%%%%%%%%%%%%%%%%%%%%%%%%%%%%%%%%%

The steady-state slip pulse solutions we employ as initial conditions have relatively long, small-amplitude tails at the healing front. An additional consequence of the computational domain periodicity in the presence of such initial conditions is that once the pulse propagates through the periodic boundary, it interacts with its own tail. Such interactions affect the dynamics of the pulse and are hence avoided, i.e., the simulations are stopped before this happens. Moreover, the initial pulse is placed such that its edge (e.g., defined by the location of its peak slip velocity) coincides with the center of the domain at $x\!=\!0$, where we also have $\tau_{\rm b}(x\!=\!0)\!=\!\tau^{\mbox{\tiny{(0)}}}_{\rm b}$. Whenever a longer evolution time is required, the domain size $W$ is increased (e.g., see results in Figs.~\ref{fig:v_snap_and_vmax}-\ref{fig:W964_trajectories}), but in general these computations are very demanding.

Finally, we note that for the constant-gradient prestress distributions $\tau_{\rm b}(x)$ there is a jump between $\tau_{\rm b}(x\!=\!-\tfrac{1}{2}W)$ and $\tau_{\rm b}(x\!=\!\tfrac{1}{2}W)$, which is formally inconsistent with the periodic boundary conditions. Yet, we verified that this jump does not affect the pulse evolution as long as its edge does not reach the domain boundary, which is indeed avoided, as explained above. Moreover, we verified that smoothly connecting $\tau_{\rm b}(x)$ across the domain boundary, to conform with the periodic boundary conditions, does not affect our results.

\subsection{Introducing initial pulse perturbations and the instability index}
\label{subsec:perturbation}

As shown in~\cite{pomyalov2023dynamics} and discussed in the manuscript, steady-state pulses under uniform prestress correspond to ``saddle configurations'' separating expanding from decaying pulses. That is, slightly perturbed steady-state pulses under uniform prestress either grow (expand) or decay (shrink), depending on the sign of the perturbation. Let us schematically denote the perturbation by $\epsilon$, to be formally defined below. In an infinite system, steady-state pulses are expected to correspond to the marginal state, $\epsilon\!=\!0$. However, fitting an infinite-system steady-state pulse into a finite, periodic domain of size $W$, as explained in Sect.~\ref{subsec:finite_periodic}, inevitably introduces perturbations (as also discussed in~\cite{pomyalov2023dynamics}). Consequently, initial pulses generated as explained in Sect.~\ref{subsec:finite_periodic} will either grow or decay under uniform prestress, i.e., they feature $\epsilon\!\ne\!0$, where the sign of $\epsilon$ determines whether the pulse grows or decays and its magnitude determines the initial growth/decay rate.

To quantify the latter, we adopt a similar perturbation scheme to that used in~\cite{pomyalov2023dynamics}. Let us denote by $\tilde{\phi}(x,t\!=\!0)$ the internal state field resulting from fitting an infinite-system steady-state pulse into a finite, periodic domain of size $W$, as discussed in Sect.~\ref{subsec:finite_periodic}. Next, we define $\epsilon$ through the following amplitude perturbation $\phi(x,t\!=\!0)\!=\!(1+\epsilon)\,\tilde{\phi}(x,t\!=\!0)$ and ask at what value of $\epsilon$, the subsequent $t\!>\!0$ dynamics of $\phi(x,t\!=\!0)$ under uniform prestress (in the periodic, finite $W$ domain) correspond to the marginal state (where decaying pulse become growing, or vice versa). We denote this value of $\epsilon$ by $\epsilon_{\rm c}$, which obviously depends on $W$ and on the level of uniform prestress $\tau^{\mbox{\tiny{(0)}}}_{\rm b}$. Specifically, for $W\!=\!480$ m, we find $\epsilon_{\rm c}\!=\!-0.4$ for $\tau^{\mbox{\tiny{(0)}}}_{\rm b}\!=\!1.03\tau_{\rm min}$ and $\epsilon_{\rm c}\!=\!-0.295$ for $\tau^{\mbox{\tiny{(0)}}}_{\rm b}\!=\!1.05\tau_{\rm min}$.

We take advantage of the above results in order to generate initial conditions using well-controlled, perturbed initial steady-state pulses pulse by varying $\epsilon$. For each uniform prestress $\tau^{\mbox{\tiny{(0)}}}_{\rm b}$, we define the instability index as $\Delta\epsilon\!\equiv\!\epsilon_{\rm c}-\epsilon$ such that $\Delta\epsilon\!>\!0$ leads to an growing/expanding pulse, while $\Delta\epsilon\!<\!0$ leads to a decaying/shrinking pulse. Furthermore, $|\Delta\epsilon|$ determines the initial rate of instability. For example, in Fig.~\ref{fig:static_stress_correciton}, we set $\Delta\epsilon\!=\!0.005\!>\!0$ under uniform prestress and indeed the $t\!>\!0$ dynamics correspond to a very slowly growing/expanding pulse. In the manuscript, we used the above procedure to generate initial pulse states characterized by $\tau^{\mbox{\tiny{(0)}}}_{\rm b}$ and $\Delta\epsilon$. These are used as initial conditions for the spatially-varying prestess problem, where a nonuniform $\tau_{\rm b}(x)$ is introduced at $t\!=\!0^+$ and the subsequent $t\!>\!0^+$ dynamics are studied, as discussed extensively in the manuscript.

\section{A\lowercase{dditional supporting results}}

Here, we present a few additional results that further support the results discussed in the manuscript.

\subsection{Dynamics in the $L\!-\!v_{\rm p}$ plane}\label{sm:vm-L}

In Figs.~\ref{fig:fig1}c,~\ref{fig:fig2}, and~\ref{fig:fig3}, we explored the reduced-dimensionality description of unsteady slip pulse dynamics in the $L\!-\!c$ plane. Specifically, we tested the degree by which the pulse dynamics under various conditions follow the special $L(c)$ line, corresponding to a family of steady-state pulse solutions parameterized by the value of the uniform prestress. In~\cite{pomyalov2023dynamics}, it was established that unsteady pulse dynamics under a given uniform prestress also follow the line corresponding to a family of steady-state pulse solutions when presented in a plane defined by other pairs of pulse properties/observables, e.g., the pulse peak slip velocity $v_{\rm p}(t)$ and its size $L(t)$.
%%%%%%%%%%%%%%%%%%%%%%%%%%%%%%%%%%%%%%%
\begin{figure*}[ht!]
\center
\includegraphics[width=0.98\textwidth]%
{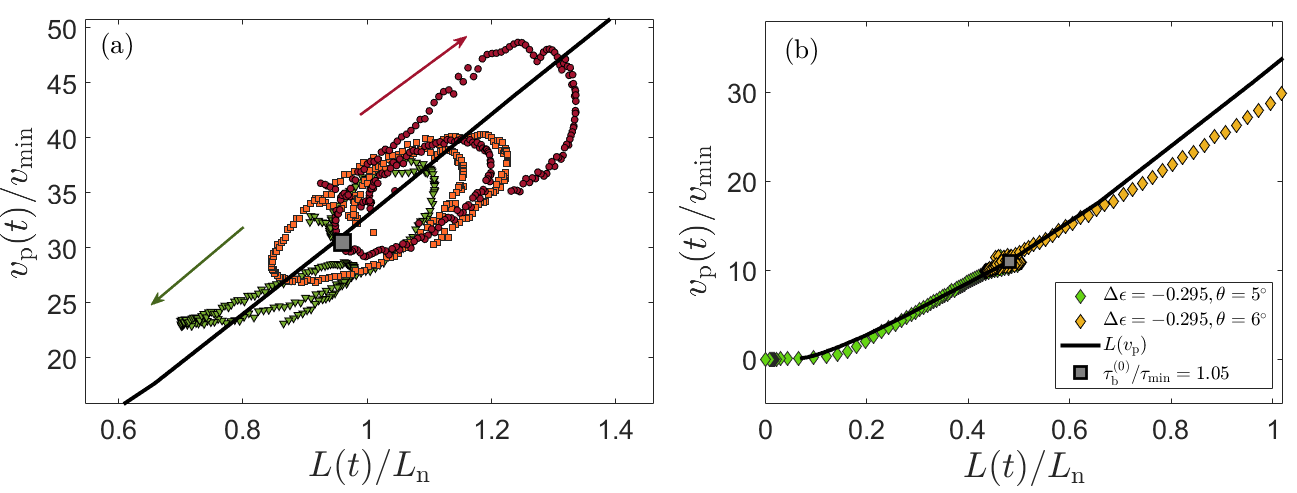}
\caption{(a) The very same pulse dynamics under periodic prestress $\tau_{\rm b}(x)$ as in Fig.~\ref{fig:fig2}c, but in the $L\!-\!v_{\rm p}$ plane (same color and symbol code as in Fig.~\ref{fig:fig2}c, see legend therein), where $v_{\rm p}$ is the pulse peak slip velocity and the solid line corresponds to the special $L(v_{\rm p})$ line. See text for discussion. (b) The very same pulse dynamics for two cases presented in Fig.~\ref{fig:fig3} under constant-gradient prestress $\tau^{\mbox{\tiny{(0)}}}_{\rm b}(x)$ (quantified by the tilt angle $\theta$, see values in the legend), but in the $L\!-\!v_{\rm p}$ plane. See text for discussion.}
\label{fig:peak_velocity}
\end{figure*}
%%%%%%%%%%%%%%%%%%%%%%%%%%%%%%%%%%%%%%%%%%%%%%%%%%%%%%%%

In Fig.~\ref{fig:peak_velocity}a, we plot the very same pulse dynamics under periodic prestress $\tau_{\rm b}(x)$ as in Fig.~\ref{fig:fig2}c, but in the $L\!-\!v_{\rm p}$ plane (same color and symbol code as in Fig.~\ref{fig:fig2}c, see legend therein), where the solid line corresponds to the special $L(v_{\rm p})$ line. It is observed that the relations between the unsteady pulse dynamics and the special $L(v_{\rm p})$ line (and the proximity to it) are very similar to the corresponding relations to the special $L(c)$ in Fig.~\ref{fig:fig2}c. Likewise, we present in Fig.~\ref{fig:peak_velocity}b the same results for two cases presented in Fig.~\ref{fig:fig3} under constant-gradient prestress $\tau^{\mbox{\tiny{(0)}}}_{\rm b}(x)$ (quantified by the tilt angle $\theta$), but in the $L\!-\!v_{\rm p}$ plane. Yet again, the results --- in particular the relations to the special $L(v_{\rm p})$ line --- are similar for the two reduced-dimensionality descriptions.
%%%%%%%%%%%%%%%%%%%%%%%%%%%%%%%%%%%%%%%
\begin{figure*}[ht!]
\center
\includegraphics[width=0.99\textwidth]{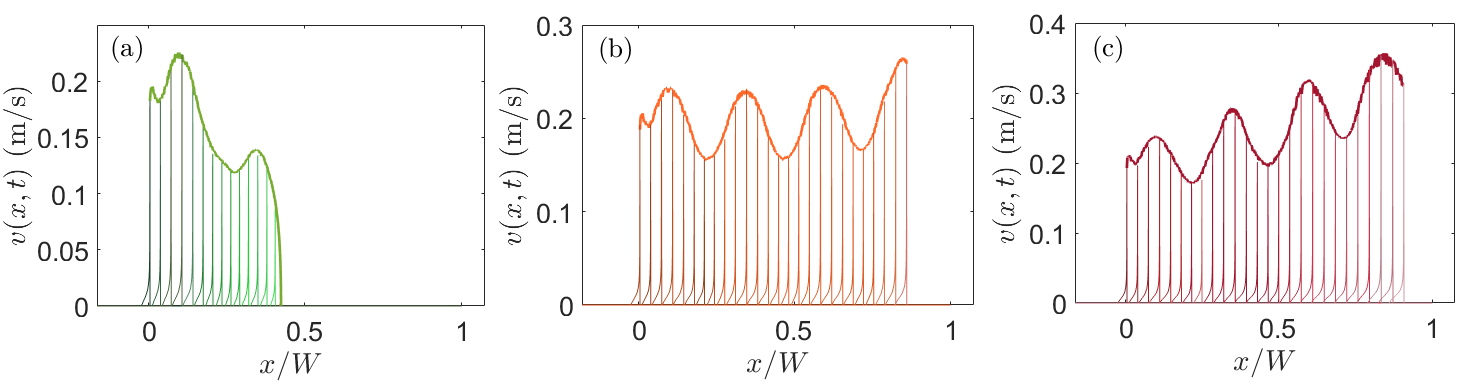}
\caption{Equal-time spatial snapshots of $v(x,t)$ of the pulse dynamics previously shown in Fig.~\ref{fig:peak_velocity}a and in Fig.~\ref{fig:fig2}c in the $L\!-\!c$ plane, but for a system size twice as large, $W\!=\!964$ m. The envelope of the peak slip velocity is added (thicker lines). (a) Results for the parameters corresponding to the green triangles in Fig.~\ref{fig:peak_velocity}a and in Fig.~\ref{fig:fig2}c (see legend therein). (b) As in panel (a), but for the parameters corresponding to the orange squares in the aforementioned figures. (c) As in panel (a), but for the parameters corresponding to the brown circles in the aforementioned figures. See the text for a discussion of the results.}
\label{fig:v_snap_and_vmax}
\end{figure*}
 %%%%%%%%%%%%%%%%%%%%%%%%%%%%%%%%%%%%%%%%%%%%%%%%%%%%%%%%

In Fig.~\ref{fig:peak_velocity} and in Figs.~\ref{fig:fig1}c,~\ref{fig:fig2}, and~\ref{fig:fig3}, we focused on the reduced-dimensionality description of unsteady pulse dynamics. In Fig.~\ref{fig:v_snap_and_vmax}, for completeness, we also present spatial snapshots of the pulse dynamics shown in Fig.~\ref{fig:peak_velocity}a and in Fig.~\ref{fig:fig2}c. We use the very same parameters for the 3 pulses presented therein (see legend of Fig.~\ref{fig:fig2}c), but employ a system size twice as large, $W\!=\!964$ m. The latter allows us to probe pulse dynamics over larger propagation distances (and hence times), and to address the nature of the emerging limit cycles for growing pulses (e.g., whether the limit cycle is approximate or exact). Presenting pulse dynamics for the larger $W$ in the reduced-dimensionality representation is, of course, possible but would be visually less clear due to the appearance of more overlapping cycles in the plane.

In Fig.~\ref{fig:v_snap_and_vmax}a, we present equal-time snapshots corresponding to the pulse dynamics presented by the green triangles in Fig.~\ref{fig:peak_velocity}a and in Fig.~\ref{fig:fig2}c. The spatial snapshots clearly demonstrate the decaying nature of this pulse, where we superimposed the envelope of the peak slip velocity (thicker line) for clarity. In Fig.~\ref{fig:v_snap_and_vmax}b, we present equal-time snapshots corresponding to the pulse dynamics presented in the orange squares in Fig.~\ref{fig:peak_velocity}a and in Fig.~\ref{fig:fig2}c. It is observed that the pulses seem to settle into a rather robust limit cycle, though there appears to be some increase in the pulse amplitude during the fourth cycle. Finally, we present in Fig.~\ref{fig:v_snap_and_vmax}c equal-time snapshots corresponding to the pulse dynamics presented in the brown circles in Fig.~\ref{fig:peak_velocity}a and in Fig.~\ref{fig:fig2}c. As is already clear from these two figures, the pulse amplitude shown in Fig.~\ref{fig:v_snap_and_vmax}c indeed systematically increases with the cycle number.

\subsection{Pulse dynamics under prestress distributions featuring smaller scale spatial variations}
\label{sm:small-lambda}

\begin{figure*}[ht!]
 \includegraphics[width=0.99\textwidth]{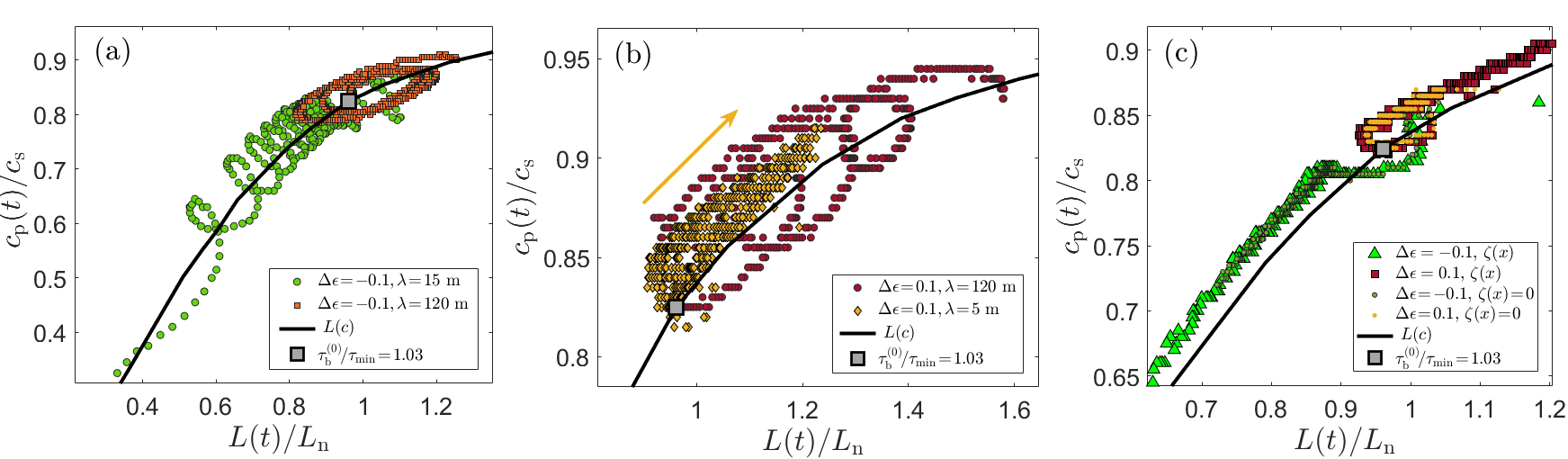}
\caption{(a) The effect of the ratio between the wavelength $\lambda$ of the periodic prestress $\tau_{\rm b}(x)$ and the size of the initial pulse $L^{\mbox{\tiny{(0)}}}\!=\!14.6$ m, with $\Delta\epsilon\!=\!-0.1\!<\!0$, $\delta\!=\!0.01$ and $W\!=\!964$ m. Qualitative differences in pulse dynamics between $\lambda\!\gg\!L^{\mbox{\tiny{(0)}}}$ and $\lambda\!\simeq\!L^{\mbox{\tiny{(0)}}}$ are observed, see legend and text for discussion. (b) Similar to panel (a), but for $\Delta\epsilon\!=\!0.1\!>\!0$, and in addition $\lambda\!<\!L^{\mbox{\tiny{(0)}}}$ is also considered, see legend and text for discussion. The yellow arrow indicates that the pulse corresponding to the yellow diamonds grows with time. (c) Pulse dynamics under prestress given by $\tau_{\rm b}(x)\!=\!\tau^{\mbox{\tiny{(0)}}}_{\rm b}\left[1+\zeta(x)\right]$, where $\zeta(x)$ is an uncorrelated noise (see text), for both $\Delta\epsilon\!>\!0$ (a growing pulse, brown squares) and $\Delta\epsilon\!<\!0$ (a decaying pulse, light-green triangles). The corresponding results for $\zeta(x)\!=\!0$, i.e., a uniform prestress, are added for comparison (see legend and text).}
 \label{fig:W964_trajectories}
\end{figure*}

Our analysis in this work focused on prestress distributions $\tau_{\rm b}(x)$ that vary over lengthscales that are larger than the size of the initial pulse $L^{\mbox{\tiny{(0)}}}$. Here, we briefly discuss pulse dynamics when this scale separation does not hold. Similarly to Fig.~\ref{fig:v_snap_and_vmax}, we consider larger systems with $W\!=\!964$ m. Since this value is different from the one used in the manuscript, to avoid any possible confusion we do not quantify the wavelength $\lambda$ of the periodic prestress $\tau_{\rm b}(x)$ through $n$, according to $n\!=\!W/\lambda$, but rather use $\lambda$ itself. That is, we rewrite Eq.~\eqref{eq:sin} as $\tau_{\rm b}(x)\!=\!\tau^{\mbox{\tiny{(0)}}}_{\rm b}\left[1+\delta\sin(2\pi x/\lambda) \right]$, making no reference to $W$. We set $\tau^{\mbox{\tiny{(0)}}}_{\rm b}\!=\!1.03\,\tau_{\rm min}$, for which $L^{\mbox{\tiny{(0)}}}\!=\!14.6$ m. In Fig.~\ref{fig:W964_trajectories}a, we first consider $\lambda\!=\!120$ m, still featuring $\lambda\!\gg\!L^{\mbox{\tiny{(0)}}}$, with $\delta\!=\!0.01$ and $\Delta\epsilon\!=\!-0.1$, which is identical to the case considered in Fig.~\ref{fig:v_snap_and_vmax}b. As observed, and similarly to other examples presented in Fig.~\ref{fig:fig2}, in this case the periodic prestress reverses the fate of the pulse, from decaying under uniform prestress ($\Delta\epsilon\!<\!0$) to a sustained pulse, nearly forming a limit cycle in the $L\!-\!c$ plane. Next, also in Fig.~\ref{fig:W964_trajectories}a, we reduced the periodic prestress wavelength to be $\lambda\!=\!15$ m, essentially identical to the initial pulse size $L^{\mbox{\tiny{(0)}}}\!=\!14.6$ m. It is observed that the lack of scale separation between the prestress spatial variation and the initial pulse size leads to a qualitative change in the pulse dynamics in this case. That is, the pulse in this case follows a spiral decay trajectory close to the special $L(c)$ line, see green circles in Fig.~\ref{fig:W964_trajectories}a.

In Fig.~\ref{fig:W964_trajectories}b, we again start with a reference case featuring $\lambda\!\gg\!L^{\mbox{\tiny{(0)}}}$, this time with with $\delta\!=\!0.01$ and $\Delta\epsilon\!=\!0.1\!>\!0$. The latter is identical to the case considered in Fig.~\ref{fig:v_snap_and_vmax}c, corresponding to a growing pulse, following upward drifting cycles in the $L\!-\!c$ plane. Next, also in Fig.~\ref{fig:W964_trajectories}b, we reduced the periodic prestress wavelength to be $\lambda\!=\!5$ m, i.e., smaller than the initial pulse size $L^{\mbox{\tiny{(0)}}}\!=\!14.6$ m. It is observed that under these conditions the pulse grows (while performing oscillations, which are not easily discerned in the figure), remaining close to the special $L(c)$ line. This behavior, up to the short wavelength oscillations, is similar to the pulse dynamics under uniform prestress (recall that $\Delta\epsilon\!=\!0.1\!>\!0$). In fact, in the limit in which the prestress varies on a spatial scale much smaller than $L^{\mbox{\tiny{(0)}}}$, we expect the pulse to smooth out the spatial variability of the prestress such that its dynamics are essentially the same for the average uniform prestress. To demonstrate this point, we set $\tau_{\rm b}(x)\!=\!\tau^{\mbox{\tiny{(0)}}}_{\rm b}\left[1+\zeta(x) \right]$, where $\zeta(x)$ is a random variable corresponding to a Gaussian distribution with zero mean and a standard deviation of $0.01$. $\zeta(x)$ features no spatial correlations, i.e., its value at each numerical grid point $x$ is independent of its values elsewhere. The results presented in Fig.~\ref{fig:W964_trajectories}c indeed demonstrate that the pulse dynamics, whether growing or decaying (see legend), smooth/average out the noise $\zeta(x)$ and identify with the corresponding uniform prestress results.
%%%%%%%%%%%%%%%%%%%%%%%%%%%%%%%%%%%%%%%%%%%%%%%%%%%%%%%%%
%%%%%%%%%%%%%%%%%%%%%%%%%%%%%%%%%%%%%%%

%\bibliography{Pulses_lib}


%\twocolumngrid

\end{document}